\documentclass[12pt,preprint]{aastex}
\usepackage[onecolumn]{emulateapj5}

\usepackage{epsfig}
\usepackage{multirow}
\usepackage{graphics}
\usepackage{amsmath, amsthm, amssymb}
\usepackage{rotating}
\tightenlines

\newcommand \lsim{\mathrel{\rlap{\lower4pt\hbox{\hskip1pt$\sim$}}
    \raise1pt\hbox{$<$}}}
\newcommand \gsim{\mathrel{\rlap{\lower4pt\hbox{\hskip1pt$\sim$}}
    \raise1pt\hbox{$>$}}}

\newcommand     \kms    {\,{\rm km~s}^{-1}}

\newcommand{\beq}{\begin{equation}}
\newcommand{\eeq}{\end{equation}}
\newcommand{\beqa}{\begin{eqnarray}}
\newcommand{\eeqa}{\end{eqnarray}}

\newcommand{\thco}	{^{13}{\rm CO}}
\newcommand{\ceto}	{{\rm C^{18}O}}

\newcommand{\Sigceto}     {\Sigma_{\rm{C18O}}}

\newcommand{\Sigsmf}     {\Sigma_{\rm SMF}}

\newcommand{\tex}	{T_{\rm ex}}
\newlength{\figwidth}
\countdef\decade=200
\advance\decade by \year
\countdef\hours=201
\advance\hours by \time
\divide\hours by 60
\countdef\mins=202
\advance\mins by \hours
\multiply\mins by 60
\multiply\hours by 100
\countdef\miltime=203
\advance\miltime by \hours
\advance\miltime by \time
\advance\miltime by -\mins

\begin{document}

\title{Mapping Large-Scale CO Depletion in a Filamentary Infrared Dark Cloud}

\author{Audra K. Hernandez}
\affil{Department of Astronomy, University of Florida, Gainesville, FL 32611, USA;\\ audrah@astro.ufl.edu}
\author{Jonathan C. Tan}
\affil{Departments of Astronomy \& Physics, University of Florida, Gainesville, FL 32611, USA;\\ jt@astro.ufl.edu}
\author{Paola Caselli}
\affil{School of Physics \& Astronomy, The University of Leeds, Leeds, LS2 9Jt, UK;\\ P.Caselli@leeds.ac.uk}
\author{Michael J. Butler}
\affil{Department of Astronomy, University of Florida, Gainesville, FL 32611, USA;\\ butler85@astro.ufl.edu}
\author{Izaskun Jim\'enez-Serra}
\affil{Harvard-Smithsonian Center for Astrophysics, 60 Garden St., 02138, Cambridge, MA, USA;\\ ijimenez-serra@cfa.harvard.edu}
\author{Francesco Fontani}
\affil{Institut de Radioastronomie Millim\'etrique, 300 rue de la Piscine, 38406 St. Martin d'Heres, France;\\ fontani@iram.fr}
\author{Peter Barnes}
\affil{Department of Astronomy, University of Florida, Gainesville, FL 32611, USA;\\ peterb@astro.ufl.edu}


\begin{abstract}
Infrared Dark Clouds (IRDCs) are cold, high mass surface density and
high density structures, likely to be representative of the initial
conditions for massive star and star cluster formation. CO emission
from IRDCs has the potential to be useful for tracing their dynamics,
but may be affected by depleted gas phase abundances due to freeze-out
onto dust grains. Here we analyze $\ceto$ $J=1 \rightarrow 0$ and $J=2
\rightarrow 1$ emission line data, taken with the IRAM 30m telescope,
of the highly filamentary IRDC G035.39.-0033. We derive the excitation
temperature as a function of position and velocity, with typical
values of $\sim 7$~K, and thus derive total mass surface densities,
$\Sigma_{\rm C18O}$, assuming standard gas phase abundances and
accounting for optical depth in the line, which can reach values of
$\sim 1$. The mass surface densities reach values of $\sim 0.07\:{\rm
  g\:cm^{-2}}$.  We compare these results to the mass surface
densities derived from mid-infrared (MIR) extinction mapping,
$\Sigsmf$, by Butler \& Tan, which are expected to be insensitive to
the dust temperatures in the cloud. With a significance of $\gtrsim
10\sigma$, we find $\Sigceto/\Sigsmf$ decreases by about a factor of 5
as $\Sigma$ increases from $\sim 0.02$ to $\sim 0.2\:{\rm
  g\:cm^{-2}}$, which we interpret as evidence for CO
depletion. Several hundred solar masses are being affected, making
this one of the most massive clouds in which CO depletion has
been observed directly. We present a map of the depletion factor in
the filament and discuss implications for the formation of the IRDC.
\end{abstract}


\section{Introduction} 

Silhouetted against the Galactic background, Infrared Dark Clouds
(IRDCs) are opaque at wavelengths $\sim$ 10 $\mu$m (P\'erault et
al. 1996; Egan et al. 1998), cold ($T < 20$~K; Carey et al. 1998;
Pillai et al. 2006), and dense ($\rm n_H \geq 10^{3} - 10^{5} cm^{3}$;
Teyssier et al. 2002; Rathborne et al. 2006; Butler \& Tan 2009,
hereafter BT09; Peretto \& Fuller 2010). They are likely to be the
precursors of massive stars and star clusters as they have similar
physical conditions, such as mass surface densities, as regions with
such star formation activity (Rathborne et al. 2006; Tan 2007; Zhang
et al. 2009; Ragan et al. 2009). CO emission
from these clouds may be useful for understanding their dynamics
(e.g. Hernandez \& Tan 2011, hereafter HT11), but could be affected by
depleted gas phase abundances due to freeze-out onto dust grains,
especially in the coldest, highest density regions.

Gas phase depletion of CO, averaged along the line of sight, has been
observed in the cold ($T\lesssim 10$~K) centers of relatively low-mass
and nearby starless cores, (e.g. Willacy et al. 1998; Caselli et
al. 1999; Kramer et al. 1999; Bergin et al. 2002; Whittet et al. 2010;
Ford \& Shirley 2011). Typically, depletion is characterized by
measuring the depletion factor, $f_D$, defined as the ratio of CO
column density {\it expected} assuming standard gas phase abundances
given the column of material observed from either the mm dust
continuum emission or near infrared (NIR) dust extinction to the {\it
  observed} CO column density (typically from $\rm C^{17}O$ or $\rm
C^{18}O$). Caselli et al. (1999) estimated the expected CO column
based on mm dust continuum emission, which has the advantage of being
able to probe to high column densities, but is sensitive to the
adopted dust temperature and emissivity. They concluded depletion
affected a region at the core center containing about $2\:M_\odot$ of
gas, where $n_{\rm H}\gtrsim 10^5\:{\rm cm^{-3}}$, with depletion
factors of up to $\sim$10 where the mass surface density is
$\Sigma\simeq 0.6\:{\rm g\:cm^{-2}}$. Kramer et al. (1999) estimated
the expected CO column based on NIR extinction, which does not require
knowing the dust temperature, but does require there to be a
sufficient areal density of background stars detectable in the NIR.
They found depletion factors of up to $\sim 2.5$ for regions with
$A_V\sim 20-30$~mag, corresponding to $\Sigma\sim 0.1-0.15\:{\rm
  g\:cm^{-2}}$.

Massive protostellar cores and clumps are typically more distant and
difficult to study, but CO depletion has been reported by Fontani et
al. (2006) from a study of 10 sources with median $f_D\simeq 3.2$ (but
a dispersion of about a factor of 10), Thomas \& Fuller (2008) from a
study of 10 sources with a mean $f_D\simeq 1.3$ and Lo et al. (2011)
from a study of 1 source with $f_D\sim 10$. These results rely on
estimates of the expected CO column density based on mm dust continuum
emission, are derived only for single pointings to the sources, and
can depend on radiative transfer modeling of the unresolved source
density and temperature structure (Thomas \& Fuller 2008; Lo et
al. 2011). Source to source comparisons are hampered by possible
isotopic abundance variations affecting these rare CO isotopologues.
The above sources already contain massive protostars, but it is not
clear if the depletion signal arises from the immediate surrounding
envelope or from nearby unresolved starless cores. Some of the massive
protostars studied produce ultra-compact \ion{H}{2} regions and
photodissociation of molecules could be occurring in localized
regions, which would mimic depletion.

We expect CO depletion to be widespread in the dense regions of IRDCs,
potentially affecting: the physical properties one derives from CO
emission; the mid and far infrared opacities of dust grains as CO ice
mantles build up; and thus the initial conditions of star and planet
formation in these regions. Individual resolved IRDCs, assumed to have
uniform isotopic abundances, may also be useful laboratories in which to
study the depletion process as a function of local gas conditions.

In this paper, we present IRAM 30m observations of $\ceto$ $J=1
\rightarrow 0$ and $J=2 \rightarrow 1$ emission from the filamentary
IRDC G035.30-00.33 (Cloud H in BT09; near kinematic distance of
$d=2.9$~kpc).  To look for evidence of depletion, the $\ceto$-derived
mass surface density, $\Sigma_{\rm C18O}$, is compared with the small
median filter (SMF) mid-infrared (MIR) extinction mapping derived mass
surface density, $\Sigsmf$ (BT09; Butler \& Tan 2011, hereafter
BT11). This work is motivated by the study of HT11, who used $\thco$
molecular line emission from the Galactic Ring Survey (GRS) to
estimate the mass surface densities of two highly filamentary IRDCs,
including Filament H. Assuming a constant value of $\tex=15$~K, HT11
found tentative evidence for CO depletion, but could not exclude the
possibility that other effects, such as systematic changes in the
excitation temperature or the contribution of high opacity cores, were
the cause of the observed decrease of $\Sigma_{\rm 13CO}/\Sigsmf$ with
increasing $\Sigma$. With our new higher-resolution, multi-transition
$\ceto$ data, we are able to exclude or mitigate these effects, as
well as resolving higher mass surface density structures to probe a
larger range of conditions where depletion may be occurring.

\section{Mass Surface Density from MIR Extinction Mapping}\label{S:SMF}

The 8~$\rm \mu m$ SMF mass surface density, $\Sigsmf$, map was derived
at 2$\arcsec$ resolution from the {\it Spitzer} IRAC band 4 (Galactic
Legacy Mid-Plane Survey Extraordinaire [GLIMPSE]; Benjamin et
al. 2003) image by comparing the observed intensity at each position
with the expected background intensity, estimated by interpolating the
intensities of surrounding nearby regions where median filter
smoothing is used to define the background model (see Figure 1a and
1b). Following BT09, a dust opacity of $\kappa_{\rm 8 \mu m}=7.5\:{\rm
  cm^{2}\:g^{-1}}$ was adopted, similar to the filter response and
background spectrum weighted mean IRAC band 4 opacity expected from
the Ossenkopf \& Henning (1994) thin ice mantle moderately coagulated
grain model with a gas-to-dust mass ratio of 156. This value is
somewhat higher than values adopted by other dust models (e.g. 125 is
used for the Weingartner \& Draine 2001), although a recent estimate
from depletion studies finds a gas-to-dust ratio of 141 (Draine 2011,
p265).  In any case, as described below, our study of CO depletion
compares relative abundances as a function of $\Sigma$ in the IRDC and
so is independent of this choice of overall normalization.

A correction for foreground emission also needs to be estimated. BT09
made this correction by estimating the amount of foreground emission
from a physical model of the Milky Way and given a measured kinematic
distance (assumed to be near) of the cloud. Battersby et al. (2010)
have pointed out an additional source of foreground from scattering in
the IRAC array. BT11 have developed a more accurate empirical method
for estimating the foreground emission, based on the presence of
independent saturated (high optical depth) cores, and here we use this
new method. For the region we analyze in this particular IRDC, the
values of $\Sigsmf$ are increased by about 10\% from those presented
by BT09.

$\Sigsmf$ in the filament is derived from comparison with adjacent
regions, which are assumed to have negligible MIR extinction. In
reality, we know from molecular line observations (e.g. $\thco$ from
the GRS analyzed by HT11), that these regions do have some material
present associated with the IRDC. We refer to this as the IRDC
``envelope''. The presence of the envelope and other systematic
uncertainties associated with estimation of the MIR background
intensity mean that $\Sigsmf$ becomes unreliable when $\lesssim
0.01\:{\rm g\:cm^{-2}}$. For our comparison with the mass surface
density derived from $\ceto$ emission, the $\Sigma_{\rm SMF}$ map is
regridded to the much lower resolution of the CO data (see below) and
all pixels with $\Sigsmf<0.01\:{\rm g\:cm^{-2}}$ are excluded from the
analysis. Methods of accounting for the envelope material are
discussed further in \S\ref{S:comparison}.

As noted by BT09, we must also account for locations of bright MIR
emission. Wherever the observed MIR intensity is greater than the
adopted background model an unphysical negative value of $\Sigma$ will
be estimated. Negative values of $\Sigma$ are allowed up to levels
comparable with the observed noise, but more extreme values, which are
mostly due to discrete MIR bright sources, have $\Sigsmf$ set to zero.
This causes an underestimation of the mass surface density in these
regions. We identify and exclude from further analysis remaining
(i.e. $\geq 0.01\:{\rm g\:cm^{-2}}$) pixels in $\Sigsmf$ map (smoothed
to the CO resolution) that have more than 20\% of their area occupied
by zero or negative values. In Figure~1b, these excluded pixels are
indicated with ``X'' and ``O'' symbols for the CO(1-0) and CO(2-1)
resolutions, respectively. Their exclusion is due either to the
presence of a MIR bright source or in regions where the background
modeling is inaccurate, which can sometimes occur near the edge of the
filament. Only a relatively small number of pixels are affected by
this exclusion.
In fact, our final results would not have varied significantly if this
exclusion had not been implemented.

\begin{sidewaysfigure}
\begin{center}$
\begin{array}{lcrr}
\includegraphics[height=2.in,trim=0 0 0 0, angle=0]{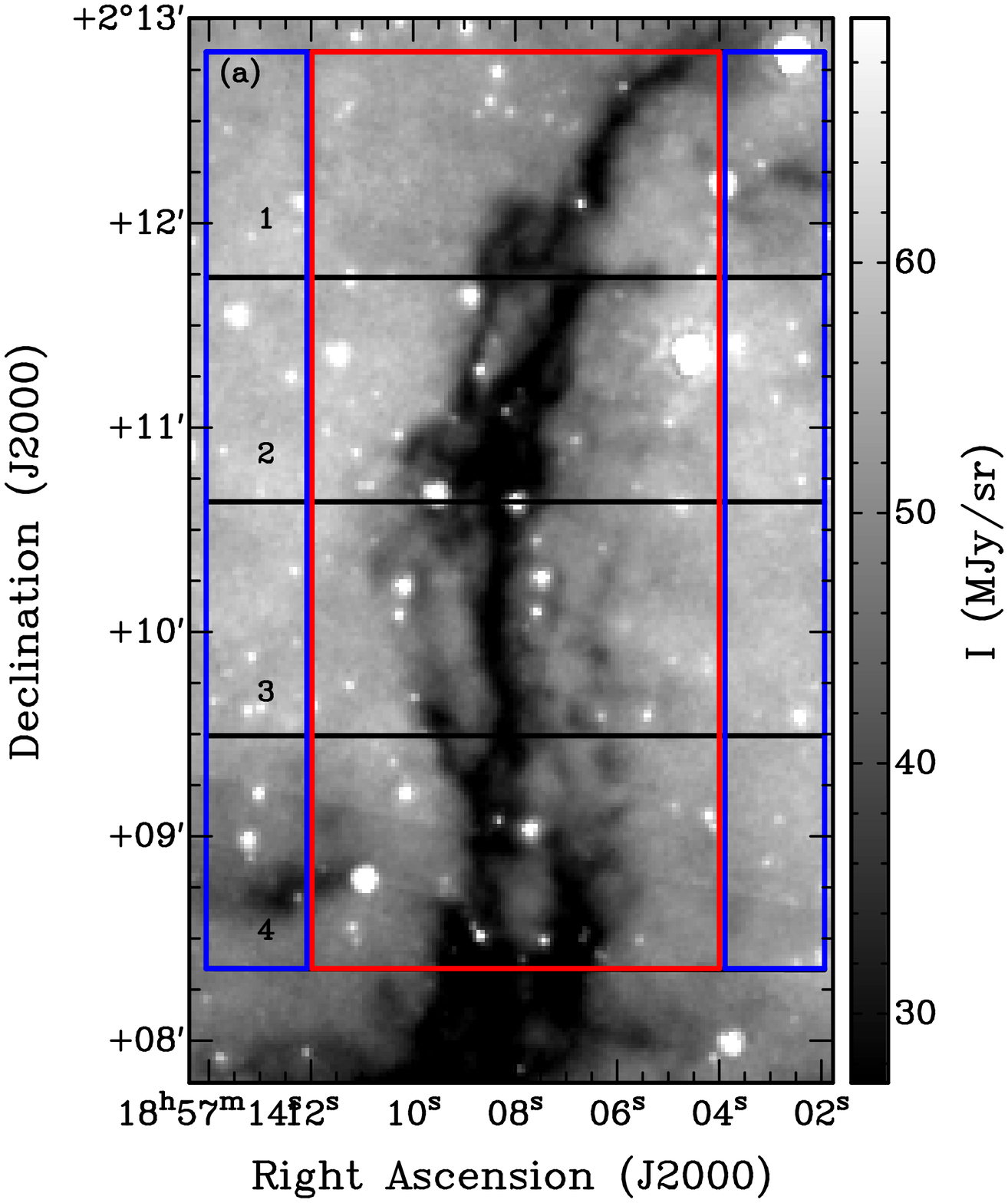} 
\includegraphics[height=2.in,trim=0 0 0 0, angle=0]{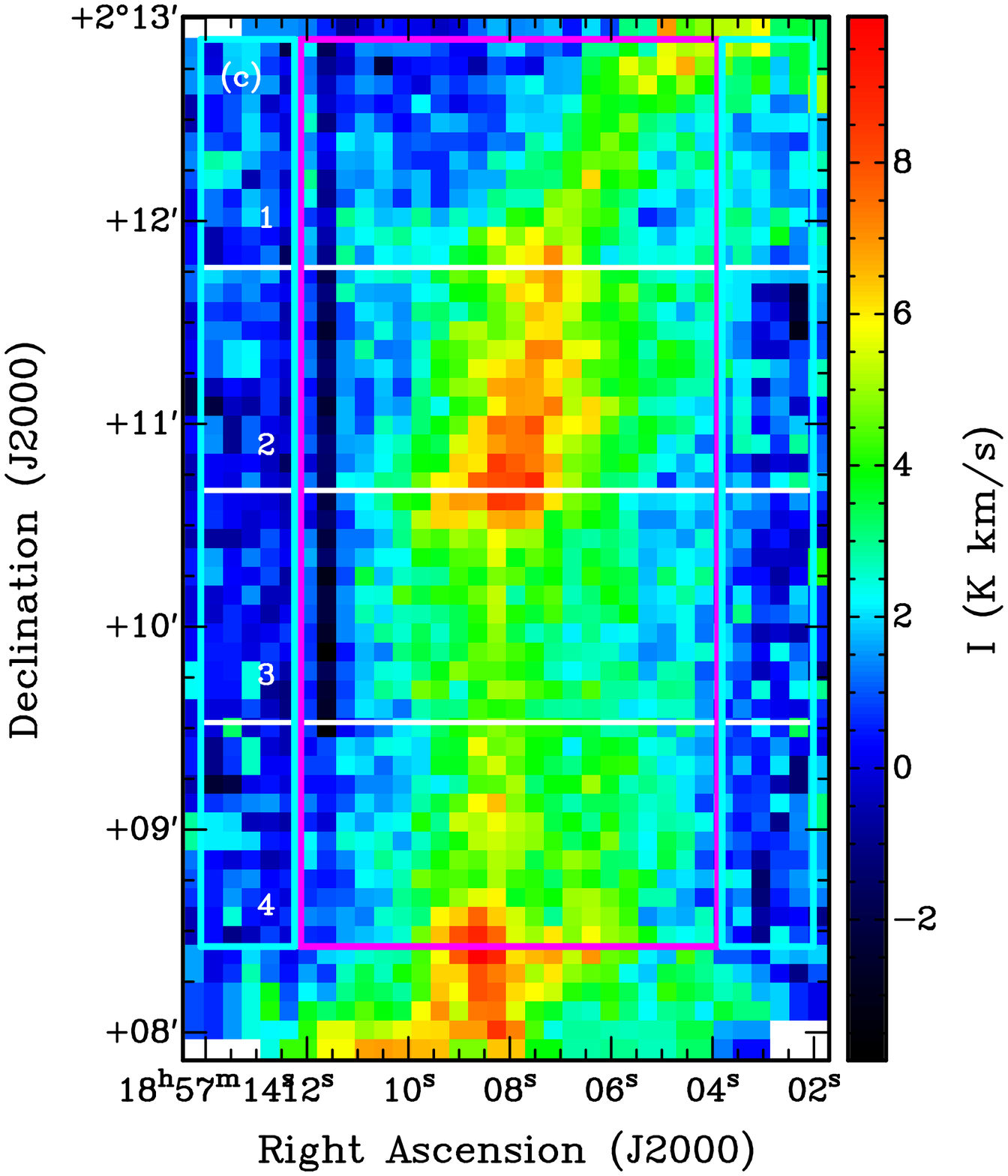} 
\includegraphics[height=2.in,trim=0 0 0 0, angle=0]{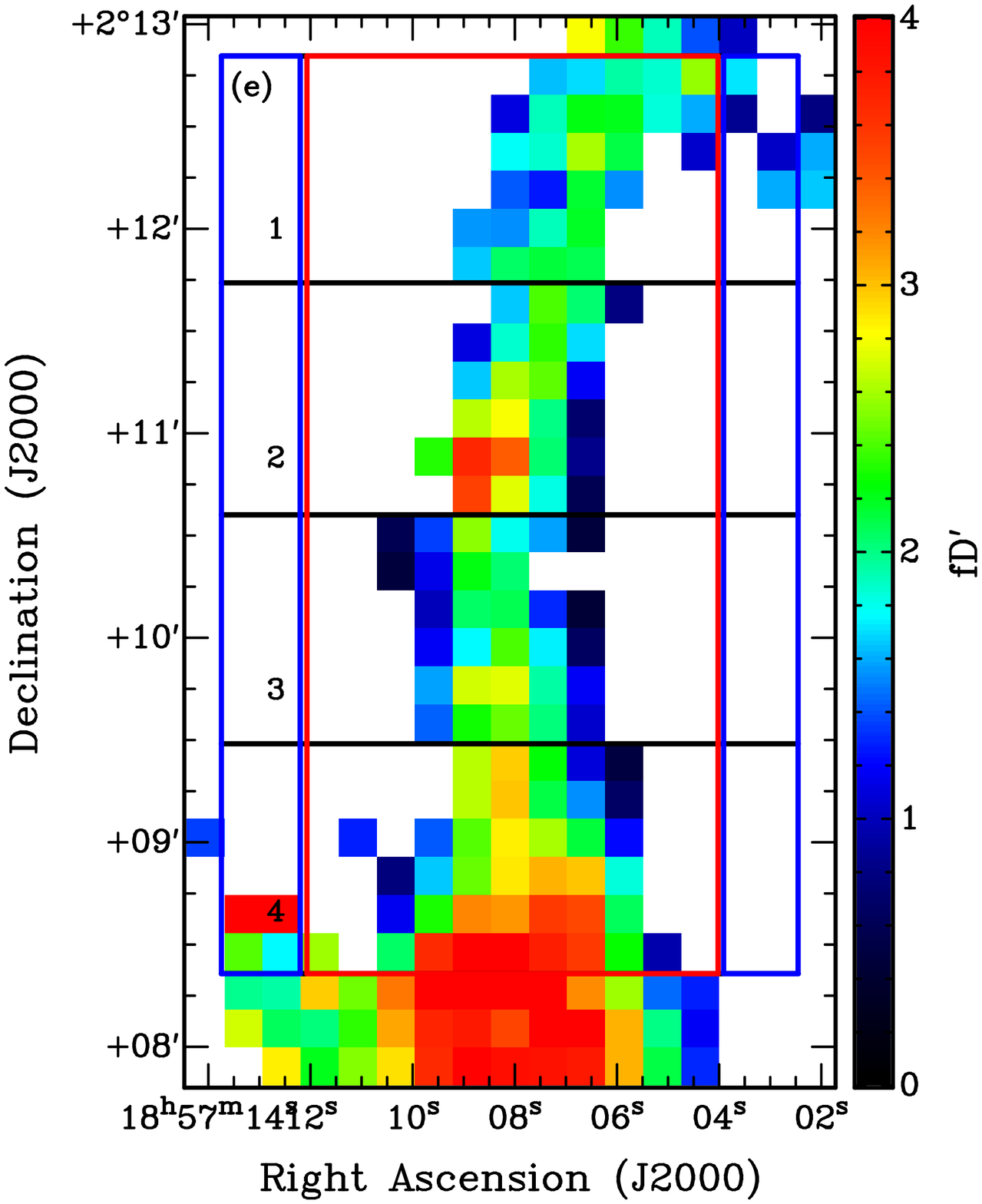} 
\includegraphics[height=2.in,trim=0 0 0 0, angle=0]{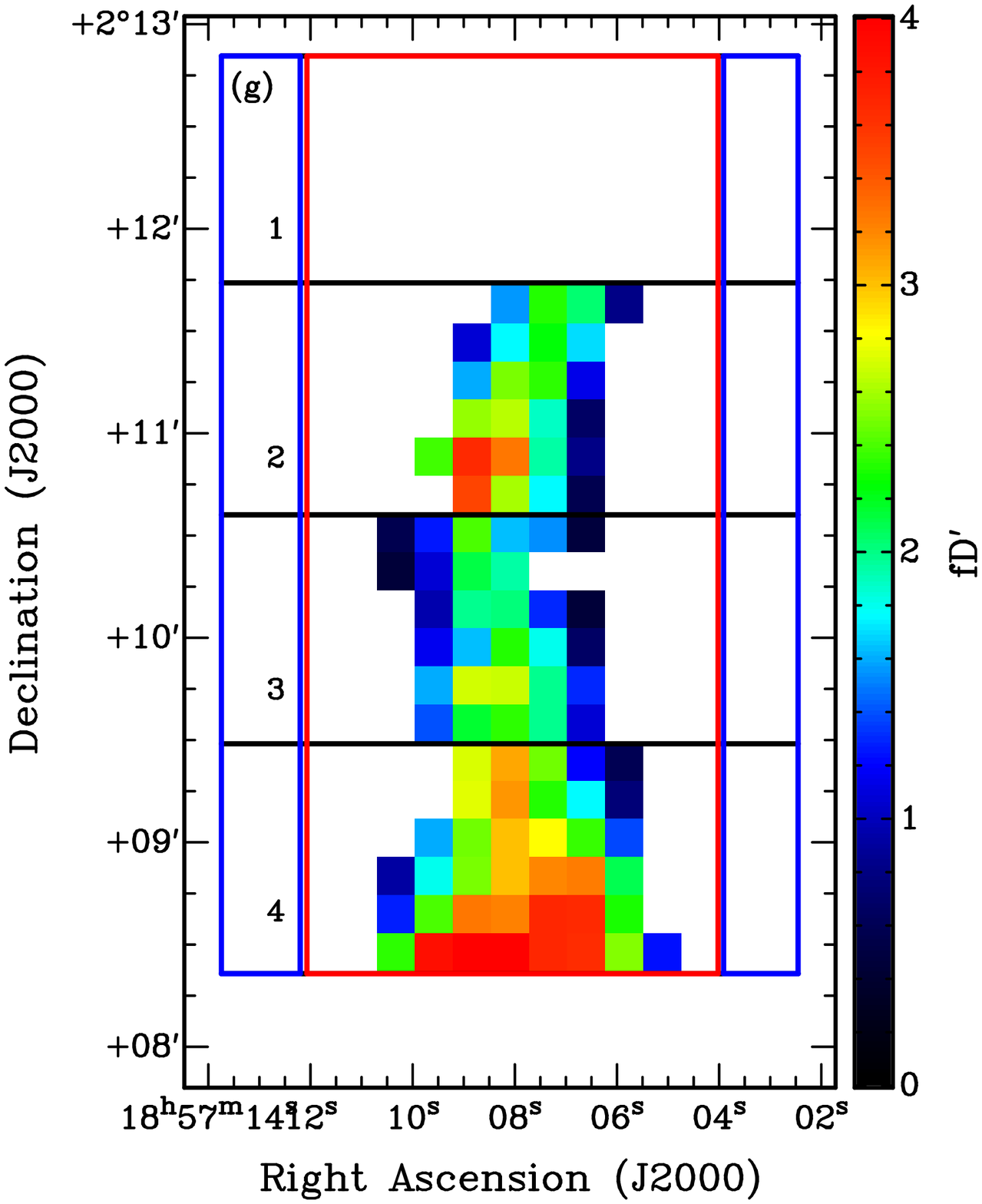} 
\\
\includegraphics[height=2.in,trim=0 0 0 0, angle=0]{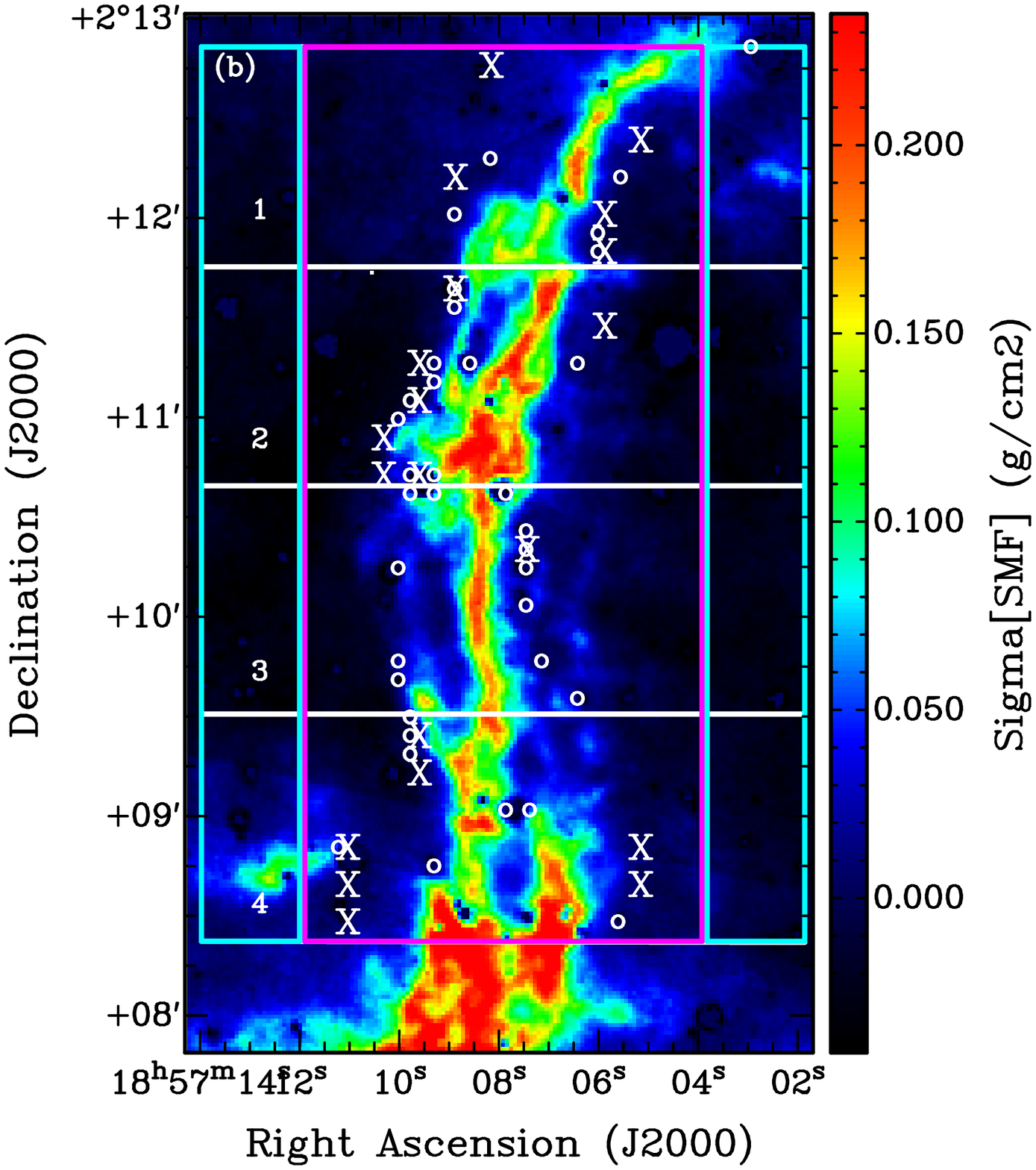}
\includegraphics[height=2.in,trim=0 0 0 0, angle=0]{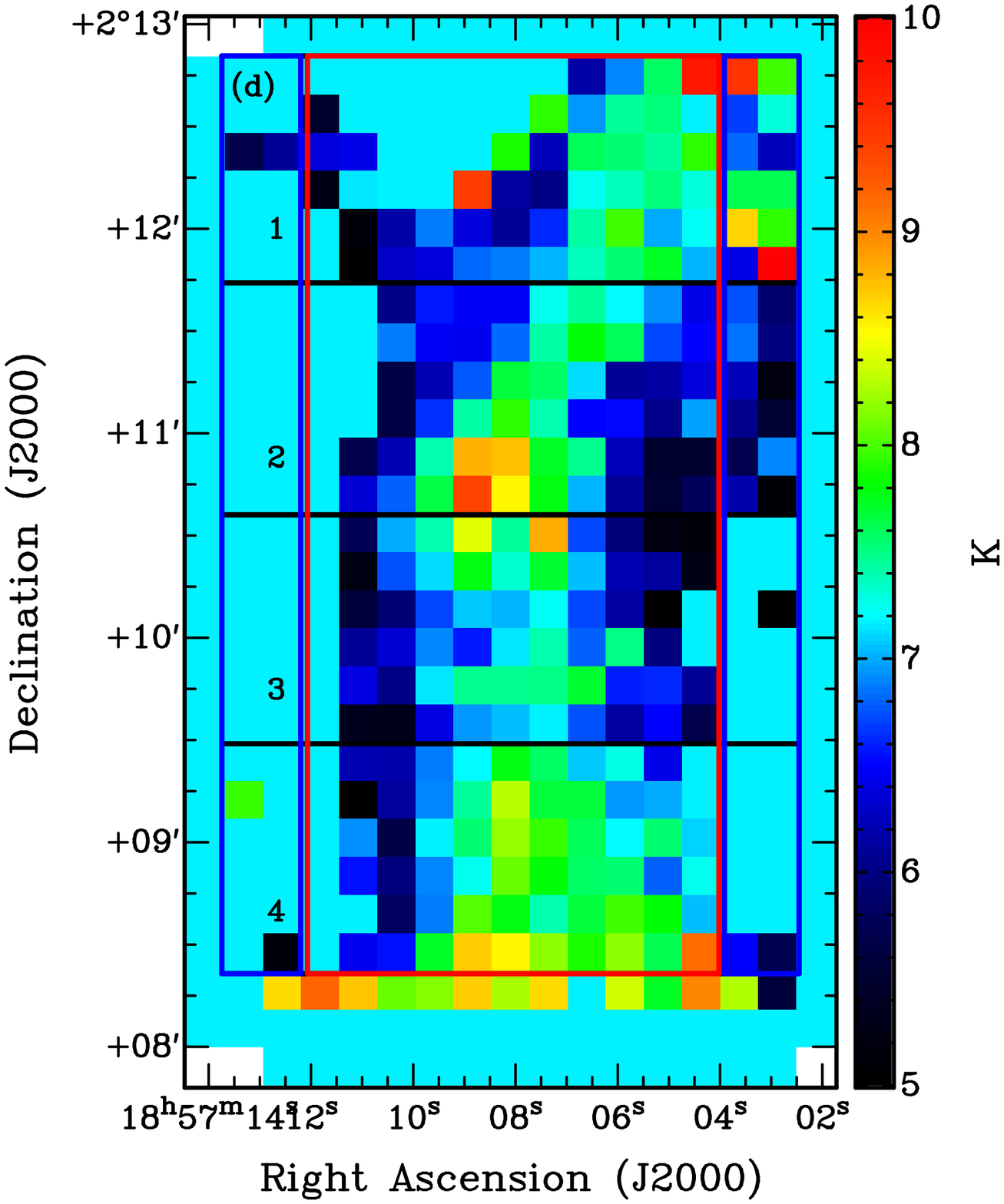}
\includegraphics[height=2.in,trim=0 0 0 0, angle=0]{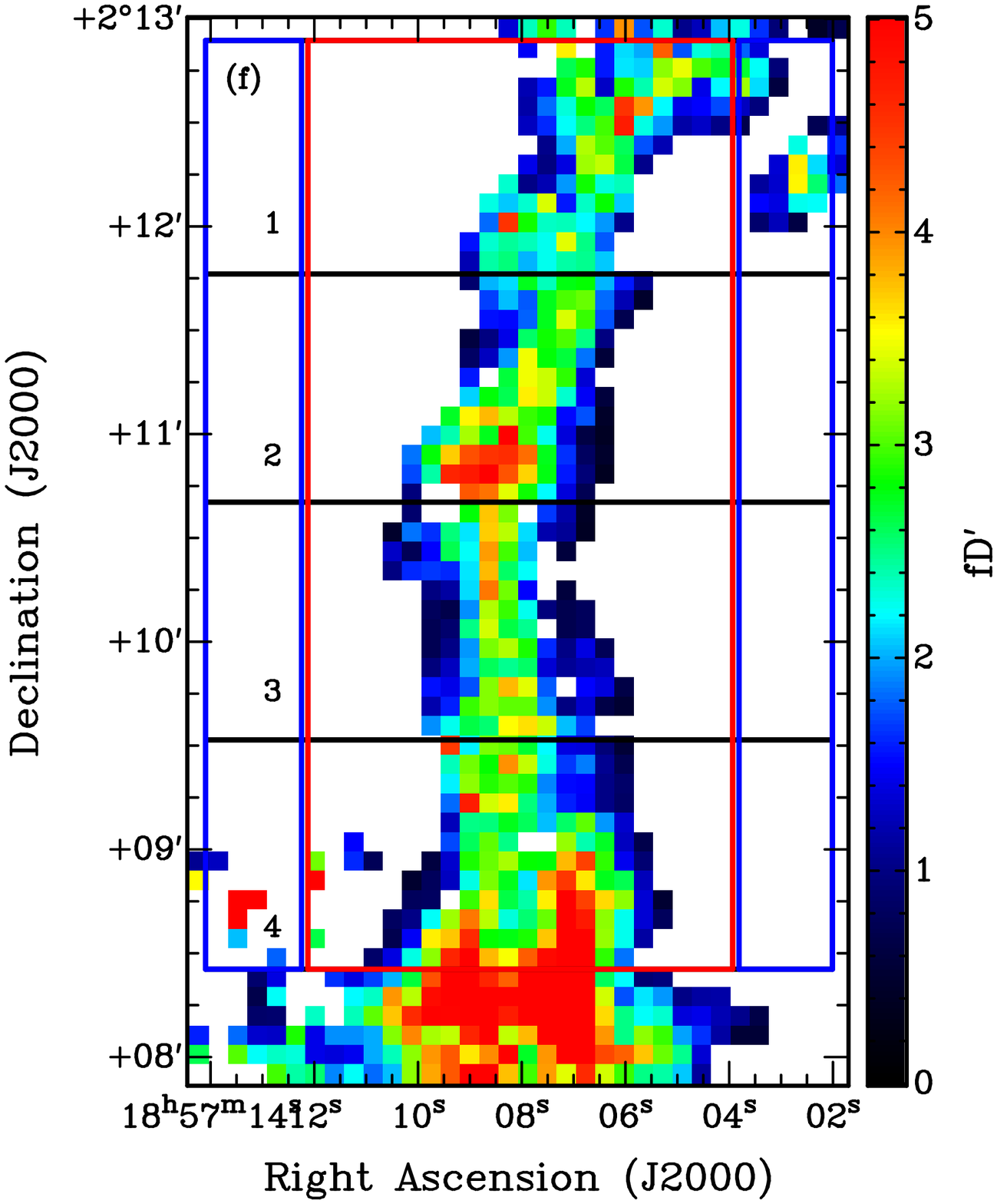}
\includegraphics[height=2.in,trim=0 0 0 0, angle=0]{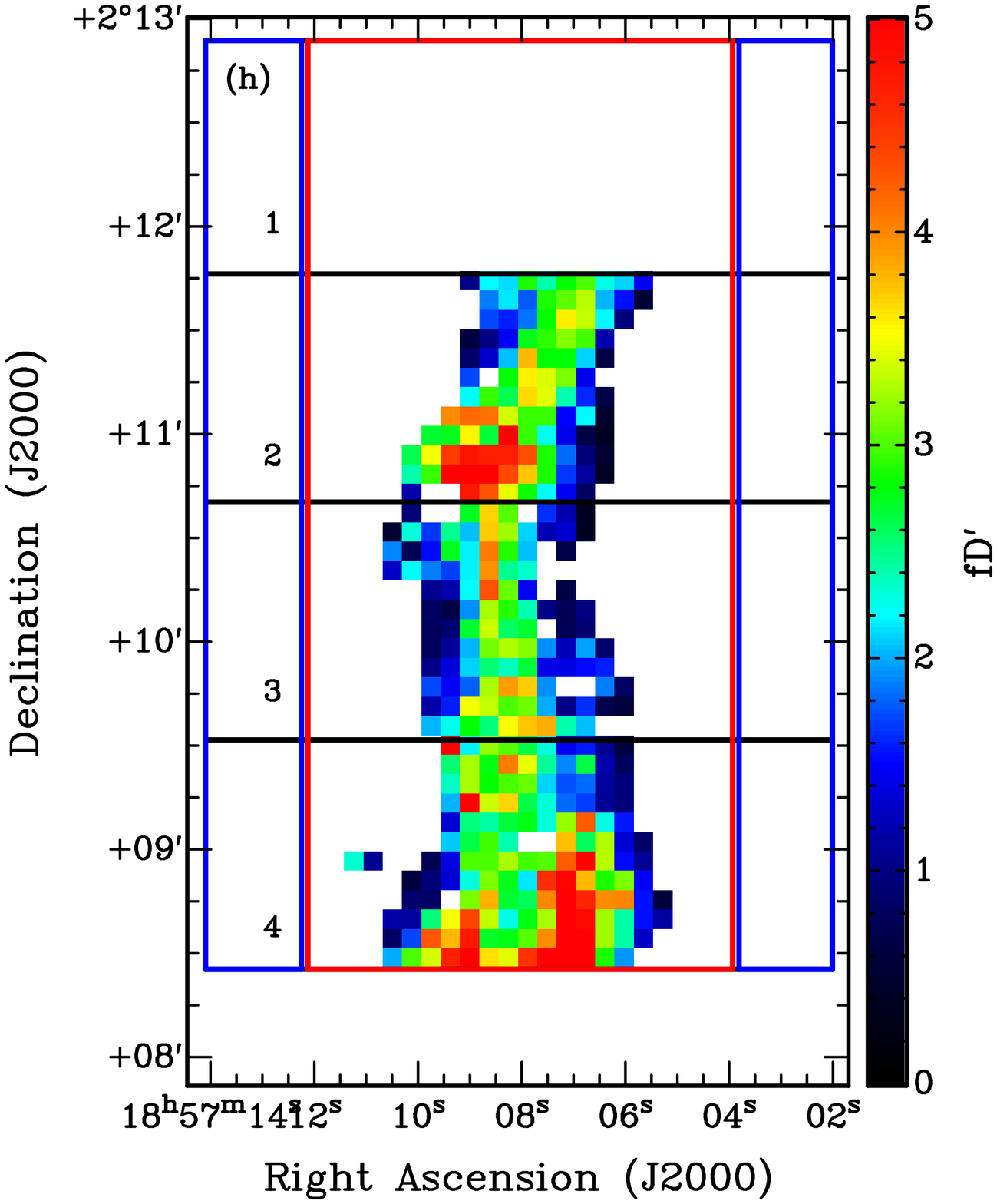}
\end{array}$
\end{center}
\caption{
\small 
Morphology and depletion maps of the IRDC.  {\it ($\rm a$) Top left:}
{\it Spitzer} GLIMPSE IRAC 8~$\mu$m image, with linear intensity scale
in MJy~$\rm Sr^{-1}$. The image has 1.2\arcsec\ pixels and the PSF has
a FWHM of 2\arcsec.  {\it (b) Bottom left:} Mass surface density,
$\Sigma_{\rm SMF}$, with linear intensity scale in $\rm g\:cm^{-2}$,
derived from the image in panel (a) using the small median filter
(SMF) MIR extinction mapping method of Butler \& Tan (2009;
2011). Regions with $\Sigma_{\rm SMF}>0.01\:{\rm g\:cm^{-2}}$ but
which are $>20\%$ affected by artifacts in the extinction map
(e.g. due to MIR bright sources) are excluded from analysis and shown
by ``X'''s and ``O'''s for CO(1-0) and (2-1) resolution grids,
respectively.
{\it (c) Top middle left:} Integrated intensity map of $\ceto$
($J=2\rightarrow 1$) emission over the velocity range of $40-50 \kms$,
i.e. the gas associated with the IRDC (HT11), in linear units of $\rm
K\:km\:s^{-1}$ and a pixel scale of 5\arcsec.  {\it (d) Bottom middle
  left:} The mean excitation temperature map weighted by the column
density in K, with pixel size of 11\arcsec.  {\it (e) Top middle
  right:} Relative depletion factor ($f_D^\prime$) map for Case 1 (no
CO envelope subtraction). {\it (f) Bottom middle
  right:} Relative depletion factor ($f_D^\prime$) map for Case 1 HiRes (no
CO envelope subtraction, $\Sigma$ derived at the CO(2-1) resolution).  
{\it (g) Top right:} Relative depletion factor
map for Case 2 (CO envelope contribution estimated via interpolation
across strips 2, 3 and 4 then subtracted; note we consider this process
unreliable for strip 1). {\it (h) Bottom right:} Relative depletion factor
map for Case 2 HiRes.
}
\label{6paneldata}
\end{sidewaysfigure}

\section{Mass Surface Density from $\ceto$ Emission}

\subsection{Observations}

The $\ceto$ $J=1 \rightarrow 0$ and $J=2 \rightarrow 1$ lines were
mapped using the IRAM (Instituto de Radioastronomia Milimetrica) 30m
antenna in Pico Veleta, Spain in August and December 2008. An area
of $2'\times4'$ was mapped using the On-The-Fly (OTF) method towards
G035.39-00.33 with a central position of $\alpha(J2000)=18^h57^m08^s$,
$\delta(J2000)=02^{\circ}10'30"$ ($l=35.517^{\circ}$,
$b=-0.274^{\circ}$). While the $\ceto$ $J=1 \rightarrow 0$ transition
was observed with the ABCD receivers with typical single side band
(SSB) rejections $>10$~dB, the $\ceto$ $J=2 \rightarrow 1$ lines
emission was mapped by using the HERA multi-beam receiver.
Off-positions for both transition lines were set to ($1830\arcsec$,$658\arcsec$).
The beam size at $\sim 110$ GHz for the $J=1 \rightarrow 0$
transitions is 22$\arcsec$, while at $\sim 220$ GHz the $J=2
\rightarrow 1$ beam size is 11$\arcsec$ .  The VESPA spectrometer
provided spectral resolutions of 20kHz and 80kHz for the $J=1
\rightarrow 0$ and $J=2 \rightarrow 1$ lines respectively, which
correspond to velocity resolutions of $\sim0.05 \kms$ and $\sim0.1
\kms$. For this study, all spectra were resampled to the same velocity
resolution of $0.2 \kms$.  The typical system temperatures were 150-220 K. 
Intensities were calibrated in units of antenna temperature (T$_{\rm A}^*$), 
and converted into a main beam brightness temperature, $\rm T_{B,\nu}$, 
via $T_A\equiv \eta f_{\rm  clump}T_{B,\nu}$, where $\eta$ is a main beam 
efficiency and $f_{\rm  clump}$ is the beam dilution factor. We use $\eta=0.64$ 
for the $J=1 \rightarrow 0$ transition, and $\eta=0.52$ for the $J=2 \rightarrow 1$
transition.  The typical 1$\sigma$ RMS noise of the data is $0.2 K \kms$ over the 
velocity range of $40-50 \kms$.  Since the $\ceto$ emission is extended over the filament,
we assume $f_{\rm clump}=1$. Figure \ref{6paneldata}c presents the
morphology of Filament H as seen in $\ceto$ $J=2 \rightarrow 1$
emission.

\subsection{Mass Surface Density and $\tex$ Estimates}

We estimate the column density of $\ceto$ molecules, ${\rm d} N_{\rm
  C18O}$, in the velocity interval ${\rm d}v$, from their emission
through the general equation:
\beq 
\frac{{\rm d}N_{\rm C18O}(v)}{{\rm d}v} = \frac{8\pi}{A \lambda_0^3} \frac{g_l}{g_u} \frac{\tau_\nu}{1-{\rm exp}\left(-h\nu/kT_{\rm ex}\right)} \frac{Q_{\rm rot}}{g_l {\rm exp}(-E_l/kT_{\rm ex})}.
\label{eq:dN}
\eeq 
Here $Q_{\rm rot}$ is the partition function for linear molecules
given by $Q_{\rm rot}= \sum_{J=0}^{\infty} (2J+1) {\rm
  exp}(-E_J/kT_{\rm ex})$ with $E_J = J(J+1)h B$, where $\rm J$ is the
rotational quantum number and B is the $\ceto$ rotational constant
equal to $5.4891\times 10^{10}\:{\rm s}^{-1}$.  $h\nu /k=5.269, 10.54$~K for 
$J=1 \rightarrow 0$ and $J=2 \rightarrow 1$ transitions, respectively.  At $\rm 7.5$~K, 
$Q_{\rm rot}=3.205$. 
$A$ is the Einstein coefficient, $6.266, 60.11 \times10^{-8}{\rm s^{-1}}$ for 
$J=1 \rightarrow 0$ and $J=2 \rightarrow 1$, respectively. $\lambda_0$ is the 
wavelength of the transition, 
$0.273, 0.137$ cm for $J=1 \rightarrow 0$ and $J=2 \rightarrow 1$, respectively. $g_l$
and $g_u$ are the statistical weights of the lower and upper levels,
and $\tau_\nu$ is the optical depth of the line at frequency $\nu$,
i.e. at velocity $v$.  The excitation temperature, $T_{\rm ex}$, is
assumed to be the same for all rotational levels. Details on the
estimation of $\tex$ are given below.  

The optical depth, $\tau_\nu$, is derived through the detection equation:
\beq
T_{B,\nu} = \frac{h\nu}{k} [f(T_{\rm ex}) - f(T_{\rm bg})] \left[1 - e^{-\tau_\nu}\right] 
\label{eq:detection}
\eeq 
where $T_{B,\nu}$ is the main beam brightness temperature at frequency $\nu$,
$f(T)\equiv [{\rm exp}(h\nu/[kT])-1]^{-1}$, and $T_{\rm bg}$ is the
background temperature of $2.725$~K. For the observable, $T_{B,\nu}$, and for an 
assumed $T_{\rm ex}$, $\tau_\nu$ can be solved for directly through 
equation (\ref{eq:detection}). Therefore, we can solve for the column 
density per unit velocity, ${\rm d}N_{\rm 18CO}/{\rm d}v$, at each 
$l,b,v$ position.

While care is taken to account for the optical depth in our column
density estimates, for reference we also state the case of the
optically thin limit of the $\ceto$ ($J=1\rightarrow 0$) column
density. If $\tau_\nu$ is small, then equation (\ref{eq:detection})
reduces to $T_{B,\nu}=(h\nu/k)[f(T_{\rm ex}) - f(T_{\rm
    bg})]\tau_\nu$. Inserting into equation (\ref{eq:dN}) gives:
\begin{eqnarray}
\frac{{\rm d}N_{\rm C18O}(v)}{{\rm d}v} & = & 6.571\times 10^{14} \frac{Q_{\rm rot}}{f(T_{\rm ex}) - f(T_{\rm bg})} [1-{\rm exp}(-h\nu/ k T_{\rm ex})]^{-1}\frac{T_A/K}{\eta f_{\rm clump}} \:{\rm cm^{-2} km^{-1}s}\\
 & \rightarrow & 9.758\times 10^{14}\frac{T_A^*/K}{\eta f_{\rm clump}} \:{\rm cm^{-2} km^{-1}s}\:\:(T_{\rm ex}=7.5~{\rm K}).
\label{eq:thin}
\end{eqnarray}
  
As in HT11, an inspection of the $\ceto$ emission in $l,b,v$ space
indicates that the gas associated with the filament is in the range of
$40-50 \kms$. The total column density per pixel is then calculated
over the entire velocity range of the filament, $N_{\rm C18O} = \int
dN_{\rm C18O}$.

The column densities for both transitions, $N_{\rm C18O}$, are
converted to a total mass surface density $\Sigma_{\rm C18O}$, by
assuming the abundance ratios of $\rm n_{16O}/n_{18O}=327$ from Wilson
\& Rood (1994) and $\rm n_{12CO}/n_{H2}=2 \times 10^{-4}$ from Lacy et
al. (1994). Thus, our assumed abundance ratio of $\ceto$ to $\rm H_2$
is $\rm 6.12 \times 10^{-7}$ and $\Sigma$ for each pixel is then
given by:
\begin{equation} \Sigma_{\rm C18O}=7.652 \times 10^{-2} \frac{N_{\rm C18O}}{10^{16}{\rm cm^{-2}}}\: {\rm g\: cm^{-2}},
\label{eq:Sigtot}
\end{equation} 
assuming a mass per H nucleus of $\mu_{\rm H}=2.34 \times 10^{-24} {\rm g}$, i.e. $\Sigma= 1\:{\rm g\:cm^{-2}}$ is equivalent to $N_{\rm H} = 4.27\times 10^{23}\:{\rm cm^{-2}}$.


In order to accurately derive the mass surface density of the
filament, an estimate of the excitation temperature, $\tex$, is
needed. To perform this estimate throughout the filament, we varied
the assumed temperature at each $l,b,v$ position until the ratio between
the column densities derived from both transitions were in
agreement.  To do this, we first defined $R_{2,1}$ as the ratio
between the $J=2 \rightarrow 1$ and $J=1 \rightarrow 0$ column 
densities:
\begin{equation} R_{2,1}\equiv \frac{{\rm d}N_{\rm C18O,21}}{{\rm d}N_{\rm C18O,10}}. 
\label{eq:Tratio}
\end{equation}
This method is similar to the one used in Kramer et al. (1999), except
they averaged over the velocity profile of their cloud.  The higher
resolution $J=2 \rightarrow 1$ data was convolved with a beam of
22$\arcsec$ and regridded to match the resolution and pixel scale of
the $J=1 \rightarrow 0$ data. For all $l,b,v$ positions above a noise
limit of $3\sigma$ in both transitions, $R_{2,1}$ was calculated first
assuming a $\tex=30$~K.  Then, $\tex$ was iteratively decreased until
$R_{2,1}$ converged to unity.  This step provided a three dimensional
grid containing estimates of $\tex$ for all positions above the noise
limit.  Next, for all positions below the noise threshold, their
$\tex$ was estimated by taking the mean excitation temperature at the
corresponding $l,b$ position.  Finally, for any remaining $l,b,v$
positions without an estimated excitation temperature, the mean $\tex$
of 7.2 K resulting from the previous steps was used.  Positions left
for this final step are mainly in the outer regions of the filament
where the emission is weak and/or the noise is high.  The column
density weighted $\tex$ map is shown in Figure \ref{6paneldata}d.

\section{Comparison of $\Sigceto$ and $\Sigsmf$:\\ Evidence for CO Depletion}\label{S:comparison}

In Figure \ref{6paneldata}, we present the morphology of the
filamentary IRDC H. The goal of this section is to compare $\Sigceto$
and $\Sigsmf$. The simplest way of doing this, which we refer to as
Case 1, involves a straightforward pixel by pixel comparison of these
values, smoothing the $\Sigsmf$ data to the resolution of the
$\ceto$(1-0) observations, for which we have derived accurate
excitation temperature information. Note, that only pixels with
$\Sigsmf$ and $\Sigceto\geq 0.01\:{\rm g\:cm^{-2}}$ are
considered. Also, pixels for which $\Sigsmf$ is affected by bright MIR
emission are excluded (see \S\ref{S:SMF}). We also perform a
comparison at the higher angular resolution of the $\ceto$(2-1)
observations, which we refer to as Case 1 HiRes, assuming $T_{\rm ex}$
at this higher angular resolution can be estimated from the values
derived at the (1-0) resolution. For both these versions of Case 1, we
refer to $\Sigceto$ as $\Sigma_{\rm C18O,TOT}$, since it is derived
from all the $\ceto$ emission associated with the IRDC and its
surrounding GMC.


However, as is apparent from Figure \ref{6paneldata}, the $\ceto$
emission is more extended than the $\rm 8\mu m$ extinction map from
Butler \& Tan (2009; 2011). This is because, as discussed above, the
extinction map is derived from an ``on-off'' comparison with adjacent
regions, which help define the background MIR intensity that is
expected to be behind the filament. Thus the MIR extinction mapping
method becomes insensitive to material present in these adjacent,
lower column density (``envelope'') regions. A fair comparison between
$\Sigsmf$ and $\Sigceto$ would allow for this envelope material. We
thus define ``filament'' and ``envelope'' regions based on the 8~$\rm
\mu m$ image of the IRDC. Following HT11, the filament is defined to
be a rectangular strip centered at $\alpha(J2000)=18^h57^m08.02^s$,
$\delta(J2000)=02^{\circ}10'35.7\arcsec$, 2.05\arcmin \ wide in R.A. and
4.47\arcmin \ long in Dec. The outline of this filament region is shown
by a red box in the panels of Figure 1. The envelope region is defined
to be made up of two adjacent rectangular regions on either side of
the filament. These are shown as blue rectangles in Figure 1 and are
each 0.56\arcmin \ wide in R.A. and 4.47\arcmin \ long in Dec. Note, that
because of the limited area mapped by our observations, these envelope
regions are narrower than those considered by HT11.

\begin{figure*}[!tb]
\begin{center}$
\begin{array}{cc}
\includegraphics[width=3in,angle=0]{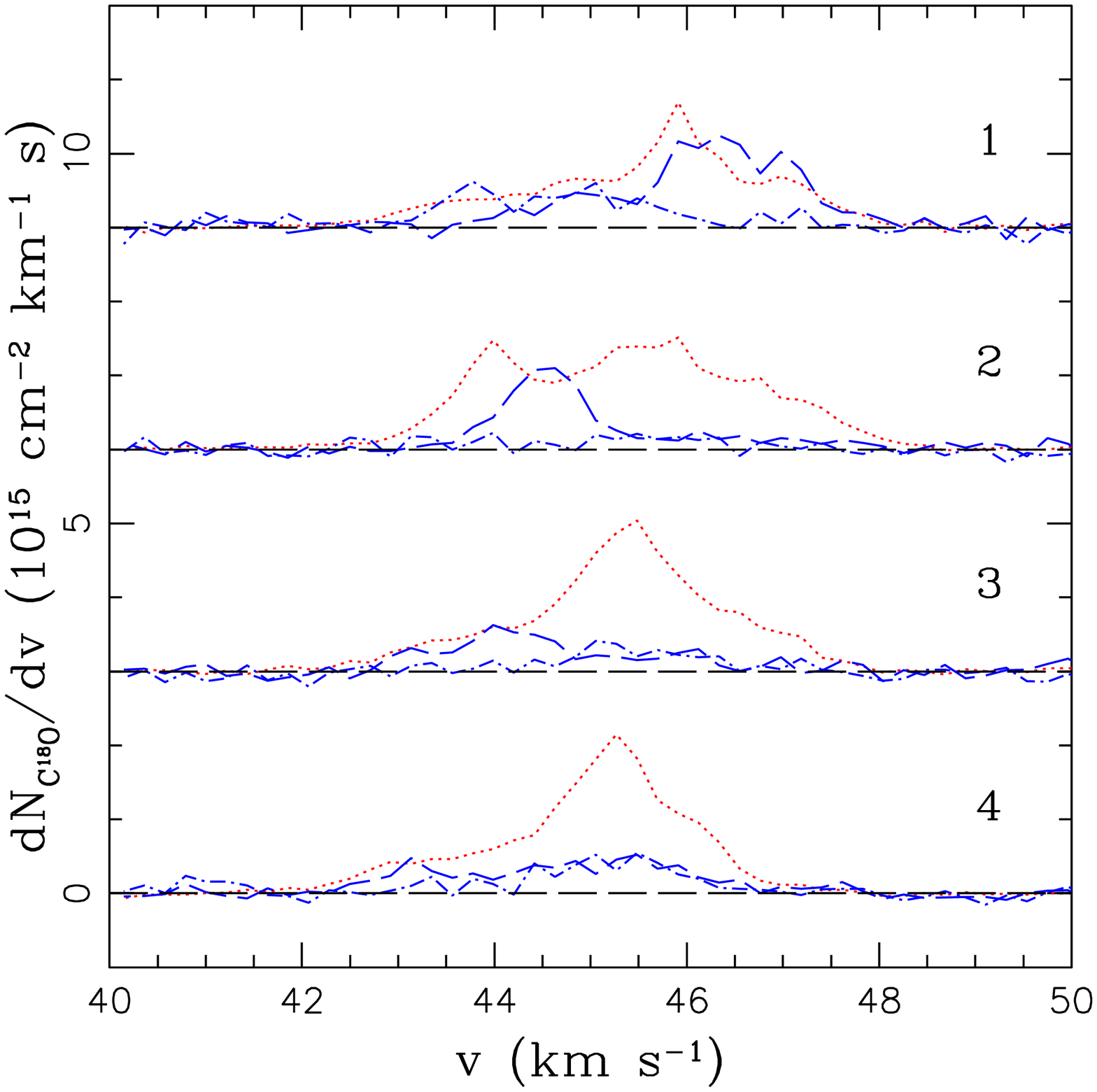} &
\includegraphics[width=3in,angle=0]{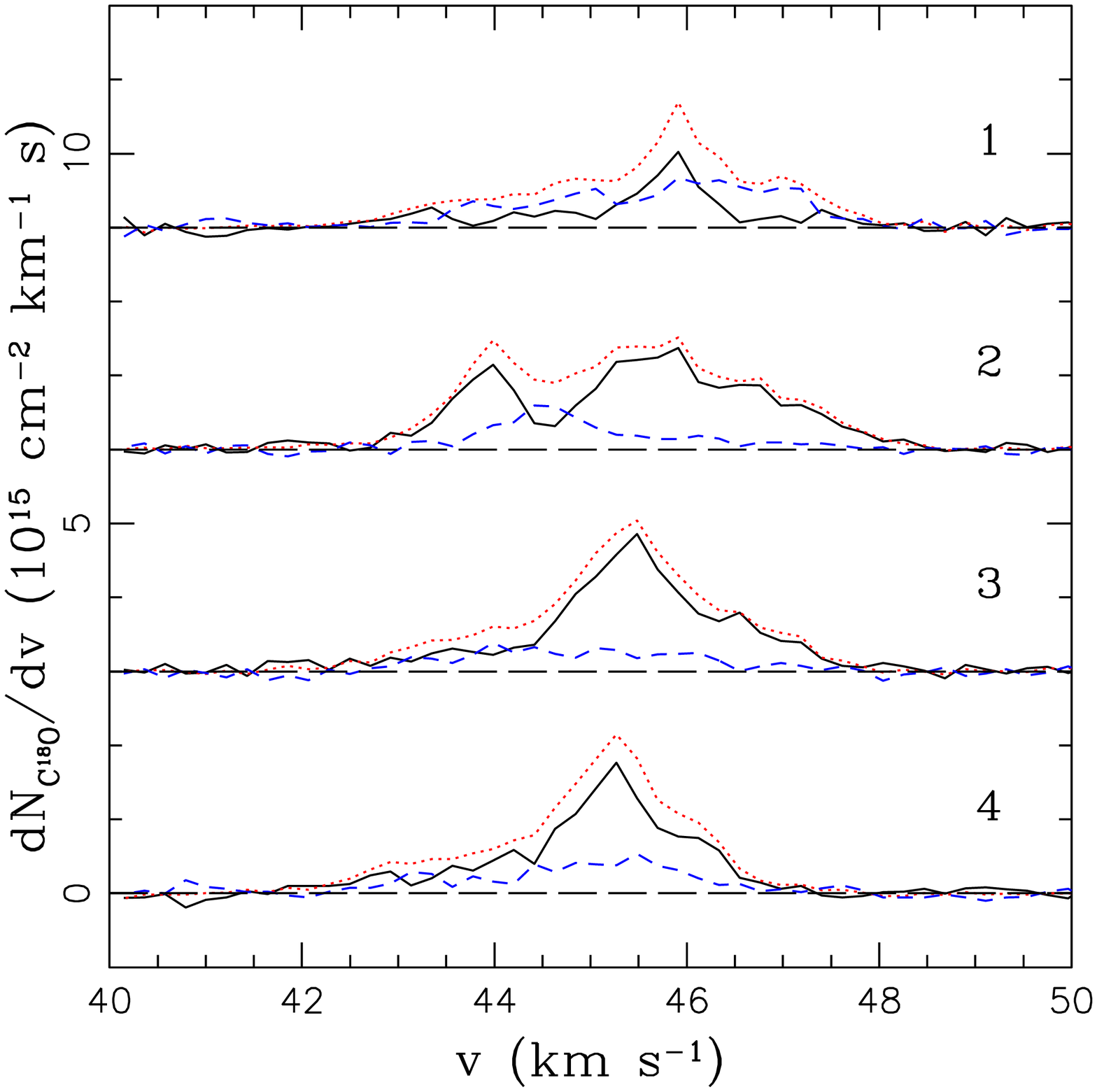} 
\end{array}$
\end{center}
\caption{
\small 
Velocity structure of the $\ceto$ molecules associated with the IRDC
and its envelope. The column density distribution, ${\rm d}N_{\rm
  C^{18}O}/{\rm d}v$, has been derived from the $\ceto$(1-0) and (2-1)
spectra, local estimates of $\tex$ and including optical depth
corrections.  {\it ($\rm a$) Left:} the 4 sets of profiles (offset to
display from top to bottom and labeled 1 to 4) correspond to the 4
strips shown in Figure~\ref{6paneldata}. The dotted, red line is the
summed contribution from gas from the central region of each strip,
corresponding to the IRDC ``filament'' (see Figure~\ref{6paneldata}
and text). The dot-dashed and long-dashed blue lines show summed
contribution from the gas from the eastern and western envelope
regions, respectively.  {\it (b) Right:} Illustration of envelope
subtraction (Case 2, see text). For the same strips as in (a), we
subtract the average of the eastern and western envelopes
(short-dashed blue lines) from the filament (dotted red lines), to
leave an estimate of the material in the filament (solid black
lines). We consider this process unreliable for strip 1, where the
envelope contains a similar amount of material as the filament.  }
\label{envspectra18}
\end{figure*}

For our Case 2, we assume that the $\ceto$ material present in the
envelope regions is also present at the similar levels towards the
filament region, and so attempt to subtract this emission from the
$\ceto$ spectrum of the filament, before then comparing to
$\Sigsmf$. To carry out this subtraction we divide the filament and
envelope into four E-W strips (1 to 4 from N to S) (see Figure 1). In
each strip, the mean column density per unit velocity is evaluated for
the filament (based on 66 $\ceto$(1-0) pixels) and the two adjacent
envelope regions (based on 18 $\ceto$(1-0) pixels each) (see Figure
\ref{envspectra18}), using the $\tex$ estimates described
previously. The envelope spectra are averaged and then subtracted from
the filament. The total column of this envelope-subtracted spectrum is
evaluated and used to derive $\Sigma_{\rm C18O,FIL}$. This is of
course an approximate method for accounting for the envelope material:
one can see from Figure \ref{envspectra18} that the envelope spectra
on either side of the filament can be quite different, especially for
strips 1 and 2. The uncertainty in the envelope-subtracted spectrum
becomes large when the envelope spectra are of similar strength as
that of the filament, as is the case for strip 1. Thus we do not
regard the results of envelope subtraction for strip 1 as being
reliable, and we exclude these pixels from the Case 2 analysis.  As
with Case 1, we also perform a Case 2 HiRes analysis, using
$\Sigma_{\rm C18O,FIL}$ estimated at the higher resolution of the
CO(2-1) data, adopting values of $\tex$ evaluated at the CO(1-0)
resolution.

\begin{figure*}[!tb]
\begin{center}$
\begin{array}{cc}
\includegraphics[width=3in]{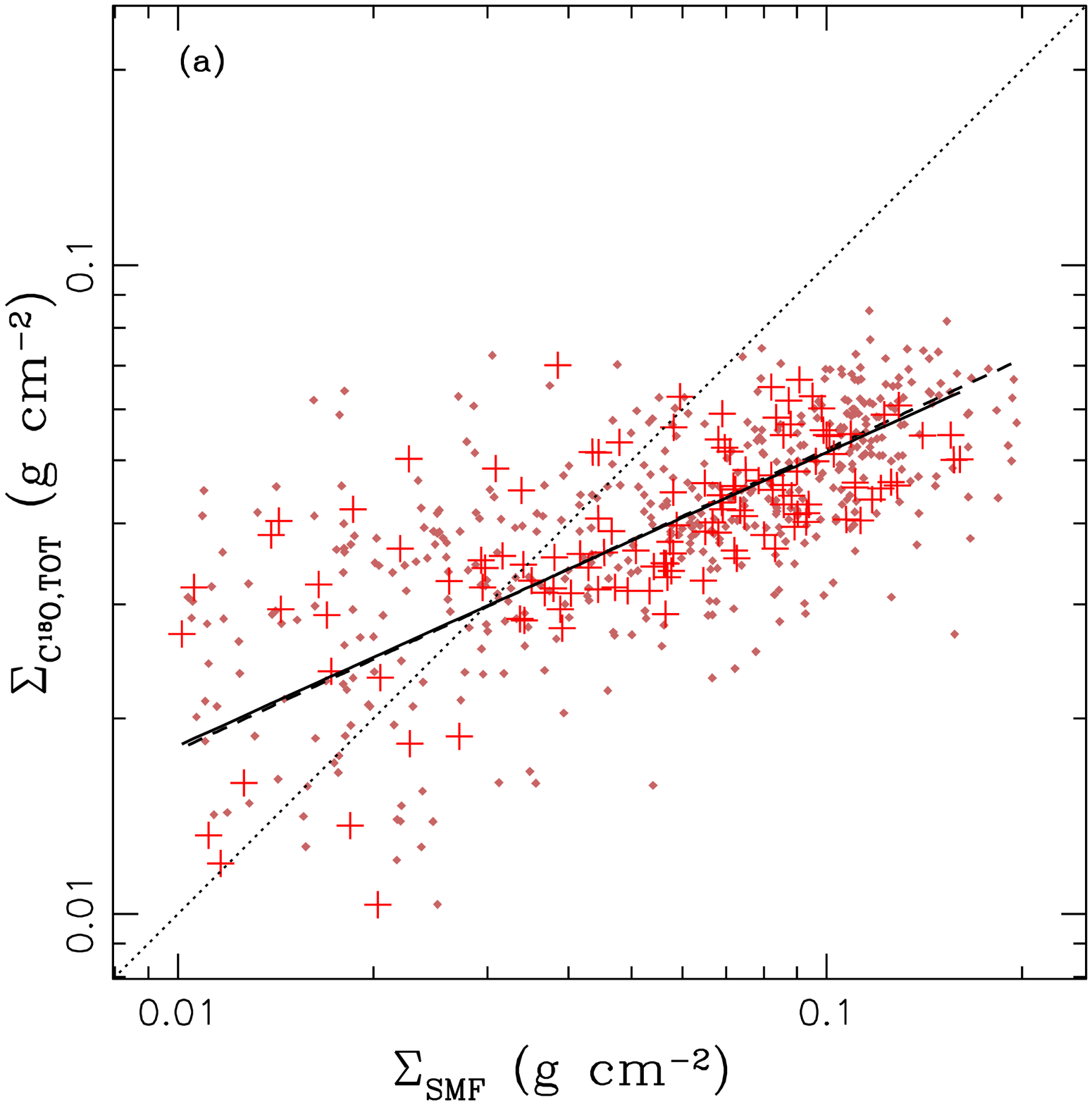} & 
\includegraphics[width=3in]{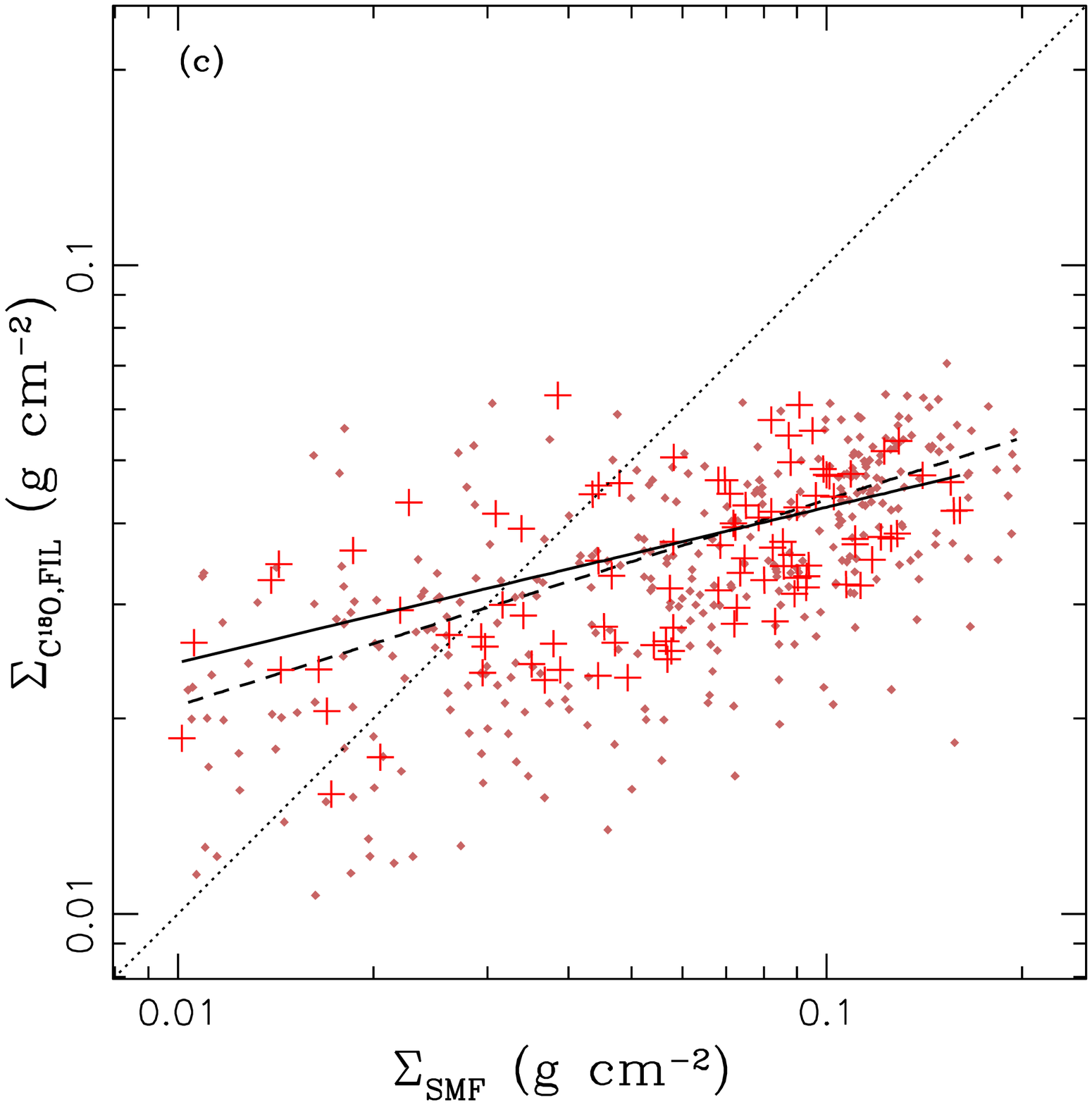}\\
\includegraphics[width=3in]{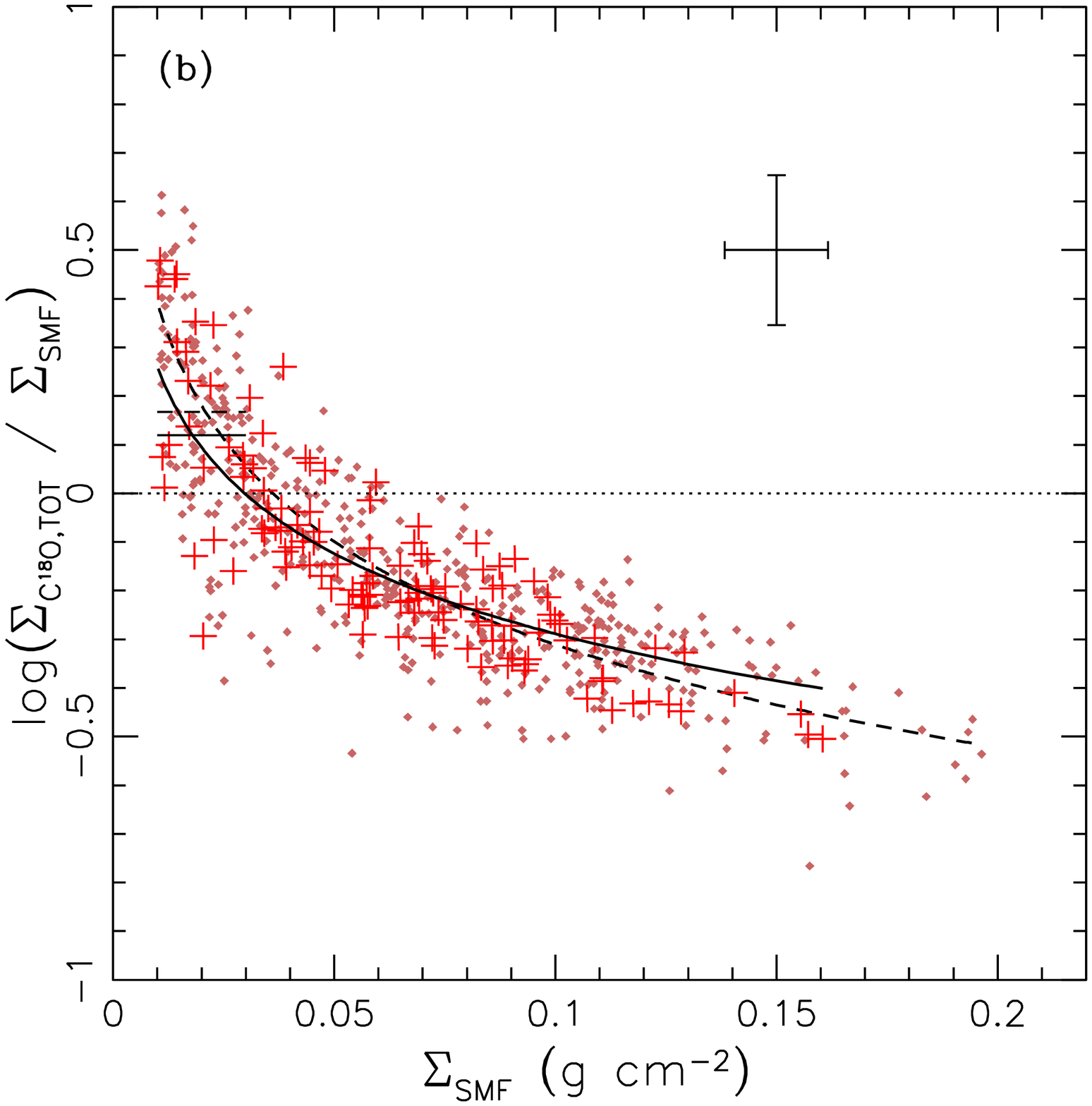} & 
\includegraphics[width=3in]{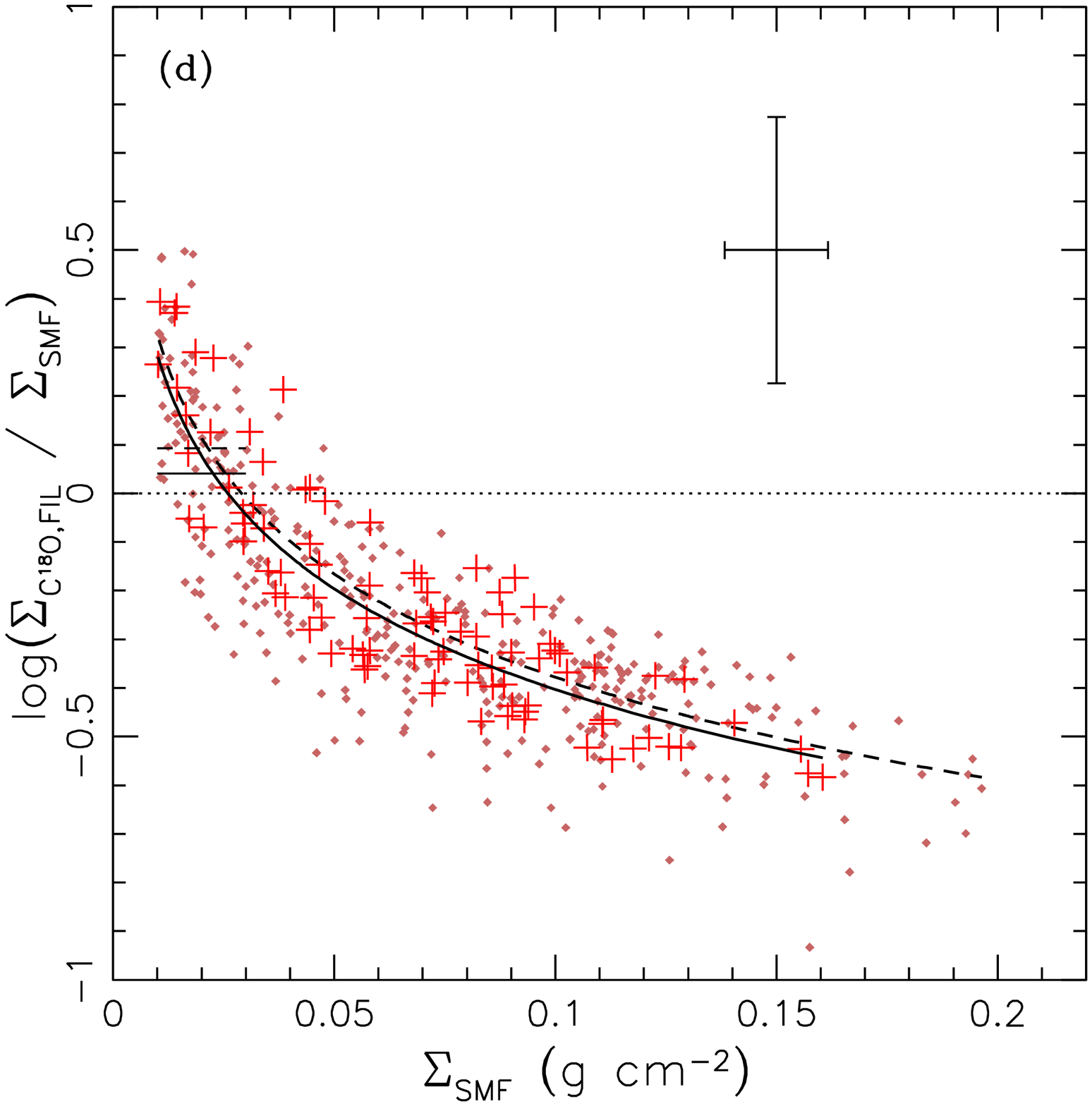}
\end{array}$
\end{center}
\caption{
\small 
Evidence for CO depletion. {\it (a) Top Left:} Comparison of
$\Sigma_{\rm C^{18}O,TOT}$ (i.e. Case 1) and $\Sigsmf$ for all
$\ceto$(1-0) (crosses) and $\ceto$(2-1) (dots) pixels for which both
$\Sigceto$ and $\Sigsmf>0.01\:{\rm g\:cm^{-2}}$ and the pixel is
$<20\%$ affected by $\Sigsmf$ artifacts, e.g. due to MIR bright
sources. The dotted line shows the condition $\Sigma_{\rm
  C^{18}O,TOT}=\Sigsmf$.  The solid, dashed lines show the best-fit
power law relations to the $\ceto$(1-0), $\ceto$(2-1) resolution data,
respectively. {\it (b) Bottom Left:} Ratio $\Sigma_{\rm
  C^{18}O,TOT}/\Sigsmf$ (i.e. Case 1) versus $\Sigsmf$, with the same
symbol and line notation as in (a). The horizontal solid, dashed lines
from $0.01<\Sigsmf/{\rm g\:cm^{-2}} <0.03$ indicate the mean values of
the data in this range for the $\ceto$(1-0), $\ceto$(2-1) resolution
data, respectively. The cross in the upper-right corner indicates
typical estimated uncertainties. {\it (c) Top Right:} Same as (a), but
now estimating $\Sigma_{\rm C^{18}O,FIL}$ from molecular gas
associated with the filament after envelope subtraction (Case 2) in
strips 2, 3 and 4. {\it (d) Bottom Right:} Same as (b), but for Case
2. Both (b) and (d) show that $\Sigceto/\Sigsmf$ decreases by up to a
factor of $\sim 5$ as $\Sigsmf$ increases from $\sim 0.02\:{\rm g\:cm^{-2}}$
up to $\sim 0.2\:{\rm g\:cm^{-2}}$.}
\label{Sigcomp18}
\end{figure*}

With these Case 1 and 2 methods, we now compare the pixel by pixel
values of $\Sigceto$ with $\Sigsmf$ derived from MIR extinction
mapping.  As noted in HT11, these measurements of $\Sigma$ are
essentially independent of cloud distance uncertainties. Figure
\ref{Sigcomp18}a presents $\Sigma_{\rm C18O,TOT}$ versus $\Sigsmf$,
i.e. Case 1 of no envelope subtraction. The best fit power law
relation to the CO(1-0) resolution data of $\Sigma_{\rm C18O,TOT}/{\rm
  g\:cm^{-2}} = A (\Sigma_{\rm SMF}/{\rm g\:cm^{-2}})^\alpha$ has
$\alpha=0.452\pm0.054$ and $A=0.146\pm0.023$. For Case 1 HiRes (i.e. at
the CO(2-1) resolution, adopting CO(1-0) resolution $\tex$ estimates) we find
$\alpha = 0.463\pm 0.025$ and $A = 0.151\pm0.010$. These results are
summarized in Table~\ref{tab:depletion}.

These uncertainties are derived assuming that the errors of each
individual measurement are as follows: 
for $\Sigceto$, a fixed value of $0.0024\:{\rm g\:cm^{-2}}$ (derived from
the 1$\sigma$ RMS noise of $\rm 0.2 K \kms$ over the velocity range of
$40-50 \kms$) and a 20\% error to account for uncertainties in $\tex$
assumed to be 1~K at the typical temperature of 7~K; for $\Sigsmf$, 
a 15\% error plus a systematic
error of $\rm 0.01 g\:cm^{-2}$ (BT09). At the resolution of the CO
pixels (11\arcsec for CO(1-0) and 5\arcsec for CO(2-1)), the $\Sigsmf$
measurements are independent, but the $\Sigceto$ results are not since
the telescope beam is about twice the pixel scale. Thus the above
quoted uncertainties of the power law fits assume, conservatively,
only 25\% of the pixels are used (although the derived values of the
parameters are based on fits to all of the pixels).

We argue below that $\Sigsmf$ is a more accurate measure of the true
mass surface density in IRDCs than $\Sigceto$, since one does not
expect large changes in MIR dust opacities in these environments,
based on the Ossenkopf \& Henning (1994) dust models. If this is true,
then if $\ceto$ were also an accurate tracer of mass surface density,
then we should see a one-to-one relation between $\Sigceto$ and
$\Sigsmf$, i.e. $\alpha\simeq 1$, even if $A$ (the value of
$\Sigceto/\Sigsmf$ when $\Sigsmf=1\:{\rm g\:cm^{-2}}$) is not exactly
unity because of systematic uncertainties in the absolute values of
$\ceto$ abundance or MIR dust opacities. We measure
$\alpha=0.452\pm0.054$ for Case 1 and $\alpha=0.463\pm0.025$ for Case 1
HiRes, which are significantly ($10\sigma$ and $21\sigma$) different
from one, and we interpret these results as being evidence for CO
depletion from the gas phase.

To illustrate that these results do not depend on the choice of dust
opacity per unit gas mass, we have repeated the analysis but with a
gas-to-dust mass ratio of 100 (rather than our fiducial value of
156). We find $\alpha=0.509\pm0.073$ (about $7\sigma$ different from
$\alpha=1$) for Case 1 and $\alpha=0.552\pm0.035$ (about $13\sigma$ different
from $\alpha=1$) for Case 1 HiRes. Note that we do not expect to derive
exactly the same values of $\alpha$ as before since we have a fixed
threshold of $\Sigma\geq 0.01\:{\rm g\:cm^{-2}}$ to include points in
the analysis and so reducing the gas-to-dust mass ratio causes us to
lose some data points near this limit.

Figure \ref{Sigcomp18}b shows the ratio $\Sigma_{\rm
  C18O,TOT}/\Sigsmf$ versus $\Sigsmf$ for our fiducial Case 1 and Case
1 HiRes analyses, with the derived power law relations overlaid. For
$0.01<\Sigsmf/{\rm g\:cm^{-2}} <0.03$ the mean values of $\Sigma_{\rm
  C18O,TOT}/\Sigsmf$ are 1.316 and 1.471 for Case 1 and Case 1 HiRes,
respectively. By the time $\Sigsmf\gtrsim 0.1\:{\rm g\:cm^{-2}}$,
$\Sigma_{\rm C18O,TOT}/\Sigsmf$ has declined to values of $\lesssim
0.4$.

In Case 2 we attempt to account for the IRDC envelope: we consider
that we can do this reliably only for strips 2, 3 and 4, where the
envelope is relatively weak compared to the filament. Figure
\ref{Sigcomp18}c presents $\Sigma_{\rm C18O,FIL}$ versus $\Sigsmf$ for
Case 2. The best fit power law relation to the CO(1-0) resolution data
of $\Sigma_{\rm C18O,FIL}/{\rm g\:cm^{-2}} = A (\Sigma_{\rm SMF}/{\rm
  g\:cm^{-2}})^\alpha$ has $\alpha=0.239\pm0.080$ and
$A=0.074\pm0.017$. For Case 2 HiRes (i.e. at the CO(2-1) resolution,
adopting CO(1-0) $\tex$ estimates) we find $\alpha = 0.317 \pm 0.038$
and $A = 0.090\pm 0.010$. These uncertainties assume the same
measurement uncertainties as Case 1, except an additional systematic
error of $\rm 0.01 g\:cm^{-2}$ has been applied to $\Sigma_{\rm
  C18O,FIL}$ due to uncertainties associated with envelope
subtraction. Again these results indicate a significant ($10\sigma$
and $18\sigma$ for Case 2 and Case 2 HiRes, respectively) departure
from a one-to-one ($\alpha=1$) relation, which we again interpret as
evidence for CO depletion. The results with a gas-to-dust mass ratio
of 100 are $\alpha=0.303\pm0.11$ (about $6\sigma$ different from
$\alpha=1$) for Case 2 and $\alpha=0.372\pm0.048$ (about $13\sigma$
different from $\alpha=1$) for Case 2 HiRes.

Figure \ref{Sigcomp18}d shows the ratio $\Sigma_{\rm
  C18O,FIL}/\Sigsmf$ versus $\Sigsmf$ for Case 2 and Case 2 HiRes,
with the above power law relations overlaid. For $0.01<\Sigsmf/{\rm
  g\:cm^{-2}} <0.03$ the mean values of $\Sigma_{\rm
  C18O,FIL}/\Sigsmf$ are 1.099 and 1.238 for Case 2 and Case 2 HiRes,
respectively. These values are smaller than their equivalents for Case
1, as is to be expected now that we are allowing for the molecular
envelope. The values are also very close to unity, suggesting that our
adopted $\ceto$ abundances and dust opacity per unit gas mass are
reasonable. Again, by the time $\Sigsmf\gtrsim 0.1\:{\rm g\:cm^{-2}}$,
$\Sigma_{\rm C18O,TOT}/\Sigsmf$ has declined to values of $\lesssim
0.4$.

\subsection{Alternatives to CO Depletion}

There are several physical processes that could be responsible for the
observed trend of decreasing $\Sigceto/\Sigsmf$ with increasing
$\Sigsmf$.  One possibility could be that our corrections for the
optical depth of the $\ceto$ emission are systematically
underestimated near the center of the filament where the column
density is large. However, the largest optical depth corrections in
the highest column density locations increase the column by only 30\%
(the highest optical depths are $\sim 1$, but lower when averaged over
the whole column), so this effect is unlikely to be driving the
observed trend.

HT11 suggested their observed trend of decreasing $\Sigma_{\rm 13CO}/\Sigsmf$
with increasing $\Sigsmf$ could potentially result if at the same time
there is a systematic decrease in the excitation temperature of about
5~K. However, from our $\tex$ estimates, we find no strong negative
temperature gradient within the IRDC towards the mass surface density
peaks. In fact, $\tex$ increases slightly towards to the center of the
filament, probably as the densities become greater than the effective
critical densities and the lower CO levels can thermalize. Thus, we
exclude trends in $\tex$ as causing the observed variation of
$\Sigceto/\Sigsmf$.

Fractionation of $\ceto$ could in principle change the local abundance
of this molecule, but the most important way in which this can be
achieved is via isotope selective photodissociation at cloud edges,
which would not be able to explain the trends of decreasing $\ceto$
abundance that we see running from $\Sigma\simeq 0.02\:{\rm
  g\:cm^{-2}}$ ($A_V\simeq 4$~mag) to $\simeq 0.2\:{\rm g\:cm^{-2}}$
($A_V\simeq 40$~mag).

Another possibility to be considered is systematic changes in $\rm
8\:\mu m$ dust opacities for gas at higher densities. If the opacity
was to increase (e.g. due to grain coagulation and/or ice mantle
formation and growth), then this could explain our observed trend of
decreasing $\Sigceto/\Sigsmf$ with increasing $\Sigma$. The Ossenkopf
\& Henning (1994) dust models do show an increase of $\kappa_{\rm 8\mu
  m}$ of 19\% going from the uncoagulated thin ice mantle model to the
uncoagulated thick ice mantle (all volatiles depleted) model. Maximal
coagulation (corresponding to that expected after $10^5$~yr at
densities of $10^8\:{\rm cm^{-3}}$or after $\sim 10^8$~yr at densities
of $\sim 10^5\:{\rm cm^{-3}}$ , which is probably more that can be
expected to have occurred since the observed densities of IRDC cores
are $\lesssim 10^{5}\:{\rm cm^{-3}}$; BT09) raises $\kappa_{\rm 8\mu
  m}$ by an additional 17\%. Thus, ice mantle growth and grain
coagulation appears to be able to account for only a small fraction of
the observed variation of $\Sigceto/\Sigsmf$.

We conclude the most likely cause of the trend of decreasing
$\Sigceto/\Sigsmf$ with increasing $\Sigsmf$ is CO depletion due to
freeze out onto dust grains. This would cause a systematic reduction
in the amount of CO gas observed in higher mass surface density
regions, which are likely to also be of higher volume
density. 

\subsection{CO Depletion and Implications}

Following the definitions of \S1 and the notation of Fontani
et al. (2006), the depletion factor is
\begin{equation}
f_D\equiv\frac{X^E_{\rm CO}}{X^O_{\rm CO}}=\frac{\Sigma_{\rm SMF}}{\Sigma_{\rm C18O}},
\label{eq:depfact}
\end{equation} 
where $X^E_{\rm CO}$ is the expected abundance of CO relative to $\rm
H_2$ given standard gas phase abundances, $X^O_{\rm CO}$ is the
observed abundance and the last equality assumes that $\Sigma_{\rm
  SMF}$ estimated from MIR extinction mapping is an accurate measure
of the true mass surface density (this assumption is discussed further
below). Given the uncertainties in the absolute values of the $\ceto$
abundance and the MIR dust opacity per unit gas mass, we renormalize
$f_D$ to be unity for the regions of the IRDC with $0.01<\Sigsmf/{\rm
  g\:cm^{-2}} <0.03$ and refer to this renormalized value as the
relative depletion factor $f_D^\prime = B f_D$, where the scaling
factor, $B=1.316, 1.471, 1.099, 1.238$ for Case 1, Case 1 HiRes, Case
2, Case 2 HiRes, respectively. We show maps of $f_D^\prime$ for these
four cases in Figure \ref{6paneldata}e-h. We note that the values of
$f_D^\prime$ presented here, peaking at values $\simeq 5$, are mass
surface density weighted averages and thus lower limits to the maximum
values of the depletion factor that apply in the densest regions of
the cloud.




We conclude that with high ($\sim 10 \sigma$) significance, widespread
CO depletion is occurring in this IRDC, with depletion factors of up
to $\sim 5$ (see Table~\ref{tab:depletion}). These values are larger
than those seen towards more evolved cores and clumps already
containing massive protostars (Fontani et al. 2006; Thomas \& Fuller
2008). Our measurement of CO depletion suffers from fewer systematic
uncertainties, especially since we do not require knowledge of the
dust temperature.

Each pixel in the lower resolution depletion maps (11\arcsec, half the
$\ceto$(1-0) angular resolution) corresponds to a length of 0.155~pc
at the cloud distance of 2.9~kpc, and so contains a mass of
$11.4(\Sigma/0.1{\rm g\:cm^{-2}})\:M_\odot$. Thus, hundreds of solar
masses appear to be affected by depletion along the filament (the
total SMF-derived mass in the 4 strips is $580\pm 230\:M_\odot$,
HT11), including a particularly prominent massive core or clump in
strip 2 and a larger clump partially in strip 4 and extending to the
south.

Thus, IRDC G035.30-00.33 is one of the most massive clouds in which CO
depletion has been detected by direct CO-based and non-CO-based
measurements of mass surface density. Our results also suggest that CO
depletion will be a common occurrence in IRDCs, since the values of
$\Sigma\sim 0.1\:{\rm g\:cm^{-2}}$ in this cloud are quite typical
(e.g. BT09). CO is therefore an imperfect tracer of a significant
fraction of the mass of IRDCs (not just the coldest, densest
cores). Accurate accounting for depletion and/or use of species
suffering minimal depletion, such as $\rm NH_3$ and $\rm N_2H^+$, are
required for more accurate dynamical studies of these clouds.

An estimate of the CO depletion timescale due to freeze-out onto dust
grains is $t_D\simeq 8000/(n_{\rm H_2,5}S)\:{\rm yr}$, where $n_{\rm
  H_2,5}$ is the number density of $\rm H_2$ molecules in units of
$10^5\:{\rm cm^{-3}}$ and $S$ is the sticking probability (of order
unity; e.g. Tielens \& Allamandola 1987) for CO on grains. We can
apply this to the thinnest region of the IRDC: the $\sim 5\arcsec$
(0.070~pc) wide filament near the center of strip 3, which appears to
have significant CO depletion with $f_D^\prime\sim 3-4$. Assuming the
depth of the filament, which has $\Sigsmf\simeq 0.2\:{\rm
  g\:cm^{-2}}$, is similar to its width, then $n_{\rm H_2,5}=2.0$ and
$t_D\simeq 4000$~yr. This provides a lower limit to the age of this
part of the IRDC. The free-fall time, $t_{\rm
  ff}=(3\pi/[32G\rho])^{1/2}$, for this density is $6.9\times
10^4\:{\rm yr}$, i.e. much longer. However, if the filament has been
created by larger scale supersonic flows, then one might expect the
high density gas to have been present for about the flow crossing time
across the width of the filament. Velocities of $\sim 10\:{\rm
  km\:s^{-1}}$ may be relevant in models of GMC-GMC collisions (Tan
2000) or if the large-scale SiO emission seen towards this filament
(Jim\'enez-Serra et al. 2010) has been created by such flows. The flow
crossing time at this speed for this part of the IRDC is only
$6800$~yr. Thus the fact that we see CO depletion in these very thin
filaments of the IRDC can help to constrain models for the cloud's
formation. For models in which the cloud lifetime is less than the
flow crossing time across the filament, a constraint is placed on the
flow speed. For the thinnest region of this IRDC, this corresponds to
flow speeds $\lesssim 17\:{\rm km\:s^{-1}}$.

\begin{table}
\centering
\caption{Parameters of Depletion Factor Analysis}
\label{tab:depletion}
\begin{tabular}{lcccc}
\hline
\hline
Case & $\alpha$ & $A$ & $B$ & $f_D^\prime$(max) \\ 
\hline
Case 1 & $0.452\pm0.054$ & $0.146\pm0.023$  & 1.316 & 3.5 (at $\Sigsmf=0.16\:{\rm g\:cm^{-2}}$) \\
Case 1 HiRes & $0.463\pm 0.025$ & $0.151\pm0.010$  & 1.471 & 4.6 (at $\Sigsmf=0.20\:{\rm g\:cm^{-2}}$) \\
Case 2 & $0.239\pm0.080$ & $0.074\pm0.017$ & 1.099 &  3.8 (at $\Sigsmf=0.16\:{\rm g\:cm^{-2}}$)\\
Case 2 HiRes & $0.317 \pm 0.038$ & $0.090\pm 0.010$ & 1.238 &  4.9 (at $\Sigsmf=0.20\:{\rm g\:cm^{-2}}$)\\
\hline	
\end{tabular}
\end{table}


\acknowledgments We thank E. van Dishoeck and R. Visser for helpful
discussions and the comments of an anonymous referee, which helped
improve the paper. AKH acknowledges support from a SEAGEP Dissertation
Fellowship. JCT acknowledges support from NSF CAREER grant
AST-0645412; NASA Astrophysics Theory and Fundamental Physics grant
ATP09-0094; NASA Astrophysics Data Analysis Program ADAP10-0110 and a
Faculty Enhancement Opportunity grant from the University of Florida.

\end{document}